\documentstyle[12pt,a41,epsfig]{article}
\def\Journal#1#2#3#4{{#1} {\bf #2}, #3 (#4)}

\def\NIMA{{\em Nucl. Instrum. Methods} A}

\def\PLB{{\em Phys. Lett.}  B}
\def\PRL{\em Phys. Rev. Lett.}
\def\PRD{{\em Phys. Rev.} D}

\def\PRP{{\em Phys. Rep. }}

\def\ASP{\em Astroparticle Physics}

\def\ra{\rightarrow}
\def\be{\begin{equation}}
\def\ee{\end{equation}}
\def\bea{\begin{eqnarray}}
\def\eea{\end{eqnarray}}

\newcommand{\lap}{\stackrel{<}{\sim}}

\newcommand{\expe}{experiment}

\newcommand{\exps}{experiments}
\newcommand{\app}{appearance}
\newcommand{\dis}{disappearance}

\newcommand{\osz}{oscillation}
\newcommand{\oszs}{oscillations}

\newcommand{\pmts}{photomultiplier}
\newcommand{\lbls}{long baseline experiments}

\newcommand{\delm}{\mbox{$\Delta m^2$}}
\newcommand{\bnel}{\mbox{$\bar{\nu}_e$}}
\newcommand{\bnmu}{\mbox{$\bar{\nu}_\mu$}}

\newcommand{\nel}{\mbox{$\nu_e$}}
\newcommand{\nmu}{\mbox{$\nu_\mu$}}
\newcommand{\ntau}{\mbox{$\nu_\tau$}}
\newcommand{\sint}{\mbox{$sin^2 2\theta$}}
\newcommand{\sk}{Super-Kamiokande}
\newcommand{\neu}{neutrino}
\newcommand{\neus}{neutrinos}
\newcommand{\cms}{\mbox{$cm^{-2}s^{-1}$}}

\newcommand{\nmune}{\mbox{$\nu_{\mu} - \nu_e$}}
\newcommand{\nmuntau}{\mbox{$\nu_{\mu} - \nu_{\tau}$}}
\newcommand{\enu}{\mbox{$E_{\nu}$}}
\begin{document}
\sloppy
\thispagestyle{empty}

\mbox{}
\vspace*{\fill}
\begin{center}
{\LARGE\bf Long baseline neutrino oscillations\footnote{Inv. Talk at Int. Workshop on Simulation and Analysis
Methods for Large Neutrino Detectors, DESY Zeuthen, 6.-9.7.1998}} \\


\vspace{2em}
\large
K. Zuber
\\
\vspace{2em}
{\it  Lehrstuhl f\"ur Exp. Physik IV, University of Dortmund,}
 \\
{\it Otto-Hahn-Str. 4, 44221 Dortmund, Germany}\\
\end{center}
\vspace*{\fill}
\begin{abstract}
\noindent
The motivation for possible future \lbls{} is discussed. The proposed \exps{} as well as
their physics potential is reviewed.
\end{abstract}
\vspace*{\fill}
\newpage
%
\section{Introduction}
\label{sect1}
A non-vanishing \neu{} rest mass has far reaching consequences from cosmology down to particle
physics \cite{kai}.
While direct \exps{} show no hints for such a mass, in the field of \neu{} \oszs{} there is growing
evidence. Beside the long standing solar \neu{} problem, in the last years growing evidence came up from
the LSND-experiment \cite{lsnd} using accelerator \neus{} and from atmospheric \neus{} especially by recent
\sk{} measurements \cite{skevi}. 
The scenarios developed to describe the observed effects include the three known \neus{} as well as possible
sterile \neus{} \cite{moh} not taking part in standard weak interaction. For a compilation of theoretical models 
see \cite{achim}.\\
In a simple two flavour mixing scheme the \osz{} probability $P$ is given by 
\be
P (E) = \sint{} sin^2(1.27 \delm L/E)
\ee
with \delm = $\mid m_2^2 -m_1^2 \mid$, L the source-detector distance and E the \neu{} energy.
The main motivation for \lbls{} is given by the chance to investigate the large mixing angle solution of the
solar
\neus{} (\nel - $\nu_X$) at \delm{} $\approx 10^{-5} eV^2$
by using nuclear reactors and to explore the atmospheric \neu{} evidence (\nmu - $\nu_X$) at \delm{} $\approx
10^{-3} eV^2$ by using
accelerators. 
The latter not only includes the proof of \nmu{} - \dis{} but includes a search for \ntau{} - \app{}.
Typical beams at accelerators are produced by protons hitting a fixed target, where the decaying secondaries (mostly
pions)
decay into \nmu{}. This dominantly \nmu{} beam is then used either for pure \nmu{} - \dis{}
searches or for \app{} searches by measuring electrons and $\tau$-leptons produced via charged current
(CC) reactions.
The \ntau{} - \app{} search requires some beam design optimisation because the exploration of low \delm{} values  
prefers lower
beam energies
but the $\tau$-production cross-section shows a threshold behaviour starting at 3.5 GeV and increasing with beam
energy (Fig. 1). A possible \osz{} of \nmu{} into sterile \neus{} might show up in the CC/NC ratio. 
Independent of the above evidences, effects of a possible CP-phase in the leptonic mixing matrix can be explored by
\lbls{} \cite{bil}.
\section{Reactor experiments}
\label{sect2}
Reactor experiments are disappearance
experiments looking for \bnel $\ra \bar {\nu}_X$.
Reactors are a source of MeV \bnel{} due to the
fission of
nuclear fuel. The main
isotopes
involved are
$^{235}$U,$^{238}$U,$^{239}$Pu and $^{241}$Pu. The neutrino rate per fission has been
measured for all isotopes except $^{238}$U and is in good agreement with
theoretical calculations.
Experiments typically
try to measure the positron spectrum, which can be deduced from the \bnel{} - spectrum,
and either compare it directly to the theoretical predictions
or measure it at several distances from the reactor and search for spectral changes.
Both types of experiments were done in the past. 
The detection reaction
is
\be
\label{gl1}
\bnel + p \ra e^+ + n
\ee
with an energy threshold of 1.804 MeV.
Normally, coincidence techniques are used between the
annihilation
photons and the neutrons
which diffuse and
thermalize within 10-100 $\mu$s. The reactions commonly used for neutron detection are $p(n,\gamma)D$ and
$Gd(n,\gamma)Gd$ resulting
in 2.2 MeV gamma photons or gammas up to 8 MeV, respectively.
The main background are cosmic ray muons producing neutrons
in the surrounding of the detector.\\
With respect to past reactor \exps{}, the current \exps{} CHOOZ and Palo Verde can already be
considered as 
\lbls{}. Their distance to the power stations of 1030 m and about 800 m respectively is already a factor of at
least three larger than any other reactor \expe{} done before. The results from CHOOZ \cite{apo98}
already exclude \nmune{} \oszs{} as explanation for the atmospheric \neus{}. Long-baseline \exps{} even by
accelerator
definitions will be done by KamLAND and BOREXINO.\\
The KamLAND \expe{} \cite{kamland} will be installed in the Kamioka mine in Japan (Fig. 2). It will contain
1000t of Liquid
Scintillator as a
main target, filled in a plastic balloon which is surrounded by buffer water with a total mass of 2500
t. At the beginning the readout will be done with 1300 20'' \pmts s corresponding to a coverage of 20
\%, an upgrade to 2000 might be possible. 
In total, there a 6 reactors with a total thermal power of 69 GW in a distance between 140 km and 210 km to
Kamioka which act as \bnel{} - sources. They produce a total \neu{} flux of $10^6$ \cms{} at Kamioka which
results for a fiducial volume of 0.5 kt and a cut on the electron energy of larger than 3 GeV in an event rate
of 250 events/year. This will allow to measure \delm{} as small as $10^{-5} eV^2$, therefore probing the large
mixing angle solution of the solar \neu{} problem. If the background can be reduced by another factor of ten
with respect to the proposed value, even the observation of solar $^{7}$Be and terrestrial \neus{} seems
feasible.\\
Originally proposed for solar \neu{} detection, also the BOREXINO \expe{} \cite{borex} has the ability to investigate
reactor
\neus{}. The \bnel-flux at Gran Sasso Laboratory is around $1.5 \cdot 10^5$ \cms{} for energies larger
than 1.8 MeV produced by power plants typically 800 km away. Without \osz{} this would result in 27
events/year in a 300 t liquid scintillation detector. The sensitivity might go down to \delm{} of $10^{-6} eV^2$
and \sint $>$ 0.2.
\section{KEK- \sk}
\label{sect3}
The first of the accelerator \lbls{} will be the KEK-E362 experiment (K2K)
\cite{keksk} in
Japan
sending a \neu{} beam from KEK to \sk. It will use two detectors, one about 300 m away from the target and
\sk{} in a distance of about 250 km. 
The \neu{} beam is produced by 12 GeV protons from the KEK-PS hitting an Al-target of 2cm $
\oslash \times$ 65 cm.
Using a decay tunnel
of 200 m and a magnetic horn system for focussing $\pi^+$ an almost pure \nmu{}-beam is produced. The
contamination of
\nel{} from $\mu$ and K-decay is of the order 1 \%. The protons are extracted in a fast extraction mode
allowing spills of a time width of 1.1 $\mu$s every 2.2 seconds. With $6 \cdot 10^{12}$ pots (protons on
target) per spill about $1 \cdot 10^{20}$ pots can be accumulated in 3 years.
The average \neu{} beam energy will be 1.4 GeV, with a peak at about 1 GeV.
The near detector (Fig. 3) consists of two parts, a 1 kt Water-Cerenkov detector and a fine grained
detector. The water
detector will be
implemented with 820 20'' PMTs and its main goal is to allow a direct comparison with \sk{} events and to study
systematic effects of this detection technique.
The fine grained detector basically consists of four parts and should provide information on the \neu{} beam profile 
as well as the energy distribution. First of all there are 20 layers of scintillating fiber
trackers intersected with water. The position resolution of the fiber sheets is about 280 $\mu$m and allows
track reconstruction of charged particles and therefore the determination of the kinematics in the \neu{}
interaction. In addition to trigger counters there is a lead-glass counter and a muon detector. The 600 lead
glass counters are used for measuring electrons and therefore to determine the \nel{} beam contamination.
The energy resolution is about 8\% /$\sqrt{E}$. The muon chambers consist of 900 drift tubes and 12 iron
plates. Muons generated in the water target via CC reactions can be reconstructed with a position resolution
of 2.2 mm. The energy resolution is about 8-10 \%.
The detection method
within \sk{} will
be identical to that of their atmospheric \neu{} detection.\\
Because of the low beam energy K2K will be able to search for \nmune{} appearance and a general \nmu{} - 
disappearance. The
main background for the search
in the electron channel might be quasielastic $\pi^0$-production in NC reactions, which can be significantly reduced
by a
cut on the electromagnetic energy. The proposed  sensitivity regions are given by \delm $> 1 \cdot 10^{-3} eV^2 
(3 \cdot 10^{-3} eV^2)$ and \sint $>$ 0.1 (0.4)  for \nmune{}(\nmuntau) \oszs{}.\\ 
The beamline should be finished by the end of 1998 and the experiment will
start data taking in 1999. In connection with the Japanese Hadron Project (JHP)  an upgrade of KEK is planned
to a 50 GeV PS, which could start producing data around
2004. The energy of a possible \neu{} beam could then be high enough to search for \ntau{} - appearance,
preferably in the
$\tau \ra \mu \nu \nu$ decay channel.\\
\section{Fermilab-Soudan}
\label{sect4}
A \neu{} program (NuMI) is also associated with the new Main Injector at Fermilab (Fig. 5). The long
baseline project will
send a \neu{} beam to the Soudan mine about 730 km away from Fermilab. Here the
MINOS experiment
\cite{minos} will be
installed (Fig. 4). It consists of a near detector located at Fermilab about 900 m away from a graphite
target
and a far detector at Soudan. The far
detector will be made of magnetized iron plates, producing a toroidal magnetic field of 1.5 T.
They have a thickness of 2.54 cm and an octagonal shape measuring 8 m across, with a transverse granularity 
of 4.1 cm. They are
interrupted by about 32000
m$^2$ active detector planes in form of plastic scintillator strips with x and y
readout to get the necessary tracking informations. 
Muons are identified as tracks transversing at
least 5 steel plates, with a small number of hits per plane. The total mass of the detector will be 8 kt.
Oscillation searches in the \nmune{} and \nmuntau{} channel can be done in several ways. The statistically most
powerful way is a \nmu{} - \dis{} search comparing the CC-rate in the near and far detector. 
Furthermore the NC/CC ratio in the far detector can be used. 
By using this ratio, information on possible \nmu -
$\nu_{sterile}$ \oszs{} can be obtained, because $\nu_{sterile}$ would not contribute to the NC rate as well. A study
of the \osz{}
parameters by itself is possible by investigating the CC and NC visible energy spectra. 
An additional hybrid emulsion detector for \ntau{} - appearance is also 
under consideration. A detector of the size 1 kt working on the same principle as OPERA (see below) would
allow a $\tau$-search on an event by event basis.
The final design of the beamline and beam energy is still under investigation depending on the physical goal
one wants to achieve.
The MINOS-project could start data taking around 2003.
\section{CERN-Gran Sasso}
\label{sect5}
A further program considered in Europe are \lbls{} using a \neu{} beam from CERN to
Gran
Sasso Laboratory \cite{ngsprop}. The distance is 732 km. The beam protons from the SPS can be extracted with energies
up to 450 GeV hitting a graphite target in a distance of 830 m to the SPS. After a magnetic horn system for focusing
a
decay
pipe of 1000 m will follow (Fig. 5).\\
Several experiments have been proposed for Gran Sasso Laboratory to do
an \osz{} search. The first proposal is the ICARUS
experiment \cite{icarus} which will be installed in Gran Sasso anyway for
the search of proton decay and solar neutrinos.
This liquid Ar TPC with a modular design, offering excellent energy and position resolution, can also be used for
long baseline
searches. A prototype of 600 t is
approved for installation which will happen in 1999. An update to 3 or 4 modules is planned.
Beside a \nmu - \dis{} search by looking for a distortion in the energy spectra,
also an \app{} search in the \nmune{} channel can be done because of the good
electron identification capabilities.
A \ntau{}-\app{} search can be obtained by using kinematical criteria as in NOMAD (Fig. 6). In
\nmu{} and \nel{} CC
events the $p_T$ distribution in the plane perpendicular to the beam is balanced between the outgoing lepton and
the hadronic final state. The angle $\Phi_{lh}$ between
the final state lepton and the hadronic final state is close to 180$^o$ and $\Phi_{mh}$, the angle
between the hadronic final state and the missing transverse momentum, is more or less uniform distributed. In case of
a \ntau{} CC event,
the undetected $\tau$ (only decay products might be observed) will balance the hadronic final state, therefore the
absolute value
of the missing $p_T$ might be larger and the value $\Phi_{mh}$ will be
oriented towards 180$^o$ because of the escaping \neus{}. For ICARUS a detailed analysis has been done for the
$\tau \ra e \nu \nu$ channel and is under investigation for other decay channels as well.\\
A second proposal (NOE)
\cite{noe} plans to build a modular detector consisting of lead-scintillating fiber and
transition radiation detectors (TRDs)
with a total mass of 6.7 kt (Fig. 7).
The calorimeter modules will be interleaved with TRDs of a total mass of 2.4 kt. 
The TRDs consist of many layers of proportional tubes and polyethylene foam as radiation material. 
One module consists of 32 layers corresponding to 8192 proportional tubes/module and additional graphite 
walls, which act as target for \neu{} interactions and sum up to 174 t/module. The TRD together with the
following calorimeter device allows particle identification as well as energy measurements of electrons,
hadrons and muons. The muon energy in the range 1-25 GeV will be determined by multiple dE/dx measurements in
the TRD. The complete detector will
have twelve modules, each 8m$\times$8m$\times$5m,
and at the end one module (muon catcher) for muon identification from interactions in the last part of the detector. 
Beside the \nmu{} - \dis{} search by measuring the NC/CC ratio this detector allows a \ntau{} - \app{} search, where
the $\tau$ decays into e,$\mu$ or $\pi$. The search criteria will be similar to the kinematic ones described
above.\\
A third proposal is the building of a 125 kt water-RICH detector (AQUA-RICH) \cite{tom}, which could be
installed outside the Gran Sasso tunnel (Fig. 7). The detector would consist of a 50m$\times$50m$\times$50m 
cube with a mirror of curvature radius 50 m and a detector plane at 27 m away from the beam entering side. The
detector
plane can be equipped with 2500 hybrid photodiodes (HPDs) allowing a 20 \% coverage and through small holes also the
mirror side 
can contain 625 HPDs corresponding to 5 \% coverage. The HPD grid would have a spacing of 1m (2m) at the detector
(mirror) plane respectively. Event parameters can be determined from the ring properties, the velocity is given
by the ring radius, the direction by the ring center and the momentum by the ring width if the width is determined 
by
multiple scattering (in contrast to normal Cerenkov detectors, where the pathlength determines the ring width).
Nevertheless the pathlength can be independently determined by the number of Cerenkov photon hits.
Furthermore the focussing of the rings allows multiple ring (n $\lap$ 4) studies. This detector allows a \nmu{} -
\dis{} search in the quasielastic $\nmu n \ra \mu p$ and resonance $\nmu p \ra \mu \Delta^{++}$ channels, a
\ntau{} - \app{} search seems feasible by looking at the $\tau \ra \mu \nu \nu$ decay channel.\\
Furthermore a \ntau{} - \app{} search with a 750t iron-emulsion sandwich detector
(OPERA) is proposed \cite{niwa}. Such a detector concept is also under consideration for MINOS as described above.
The principle idea is to use iron as a massive target for \neu{} interactions and thin emulsion sheets
conceptually working as emulsion cloud chambers (ECC) (Fig. 8).
The detector could consist of 92 modules, each would have
a dimension orthogonal to the beam of
3$\times$3 m$^2$ and in total 30 sandwiches. One sandwich is
composed out of 1 mm iron, followed by two 50 $\mu$m
emulsion sheets, spaced by 100 $\mu$m. After a gap of 2.5 mm, which could be
filled by low density material, two additional
emulsion sheets are installed. Following such a module, electronic tracking devices will be installed
for accurate extrapolation of tracks back into the emulsions.
The scanning of the emulsions is done by high speed automatic CCD microscopes.
The $\tau$, produced by CC reactions in the iron, decays in the gap region,
and the emulsion sheets are used to verify the kink of the decay. Besides the $\tau \ra e,\mu{}, \pi{}$ decay modes
also three pion decays can be examined. The analysis here is done on an event by event basis. 
The tracking devices might also allow a \nmune{} \osz{} search.\\
Finally, there is a proposal (NICE) \cite{nice} for a 10 kt iron-scintillator calorimeter, surrounded by a magnetized
iron
spectrometer. The modular design would be done with long iron and scintillator bars, dimensions of 12 m
$\times$ 2 cm$\times$ 2 cm seem feasible, building up a total detector of 12 m$\times$ 12 m$\times$ 8 m.
The scintillator is read out by wavelength shifting thin fibers, which could be coupled to devices like HPDs.
The \osz{} searches focus on the measurement of the energy spectra for \nmu{} - \dis{} searches and on measuring the
CC/NC ratio.
Ideas to merge the OPERA and NICE concepts are under consideration. 
\section{Very future projects}
\label{sect6}
A project in the very far future could be \osz{} \exps{} involving a
$\mu^+ \mu^-$-collider currently under investigation. The
created \neu{} beam is basically free of \ntau{} and can be precisely determined to be 50 \%
$\nmu (\bar{\nu}_\mu)$ and 50\% \bnel ($\nu_e$) for $\mu^- (\mu^+)$.
The collider could be constructed out of 2 straight sections connected by two arcs, where the straight regions
are used as decay regions of the muons, producing a \neu{} beam in the corresponding direction.
Because the $\mu^+ \mu^-$-collider would
be a high luminosity machine there could be a production rate for \neus{} of 10$^{20}$/year. Thus 
one even can envisage very \lbls{}, e.g. from Fermilab to Gran
Sasso with a distance of 9900 km \cite{gee97}. The interaction rate in a far 10 kt detector would be of the
order 600 \bnmu{} per year and 1000 \nel{} per year, assuming the decay of $\mu^+$ is used for the beam.\\
An even more ambitious idea is to use detectors designed for very high energy \neu{} astrophysics.
Several astrophysical sources are considered for \neu{} production even beyond 1 PeV. At these energies
the interaction length for \neus{} becomes less than the diameter of the earth. While \nmu{} and \nel{}
entering the earth from the far side with respect to a detector will be damped, the \ntau{} flux remains 
flat because with every $\tau${} - production a new \ntau{} will be created. This leads to a more or less
constant zenith angle dependence for \ntau{}. Some discovery potential might exist
with the AMANDA or planned ICECUBE detector, which could investigate very small \delm{} because of the long
baseline due to the distance of the astrophysical sources.
\section{Summary and conclusions}
\label{sect7}
The present evidences for \neu{} \oszs{} and their description in theoretical models requires
a variety of new \exps{} for confirmation. Several proposals for accelerator based \lbls{} are available
to explore the atmospheric region of evidence. In their present design most of them will fail to investigate the
complete \delm{} region given by \sk{} especially at low \delm{} (Fig. 9). Furthermore nuclear reactors
will allow to probe
one of the solutions to the solar \neu{} problem directly. Together with running and planned short baseline \exps{}
it might be possible to merge into a coherent picture of \neu{} mass models in the future.

\newpage
\begin{center}

\begin{tabular}{cc}
\mbox{\epsfig{file=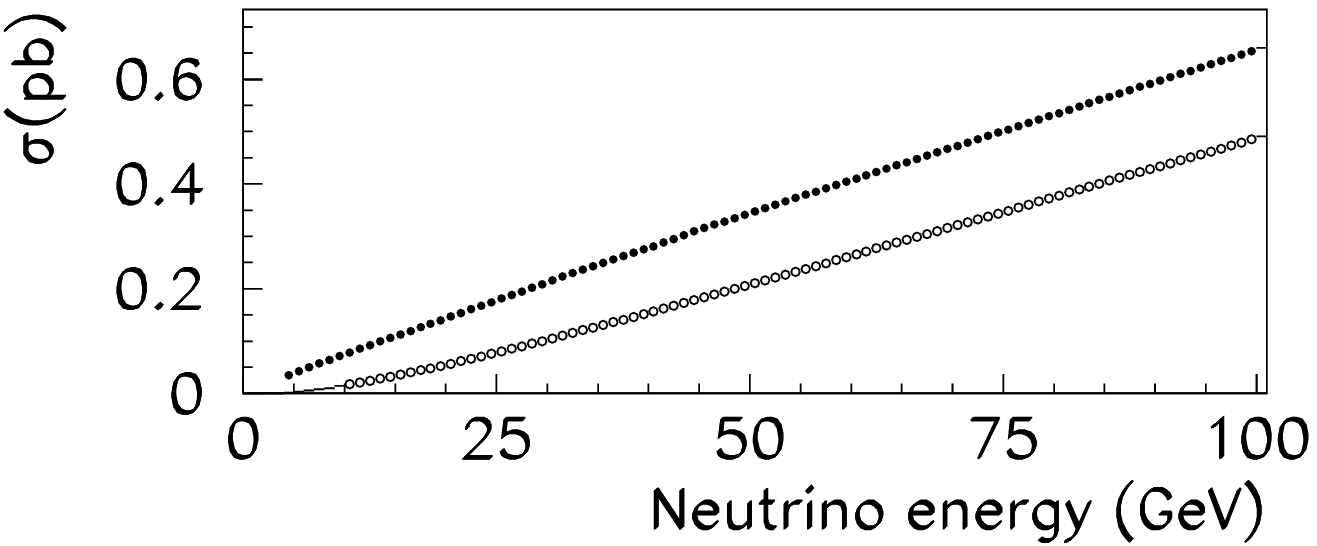,height=5cm,width=7cm}} &
\mbox{\epsfig{file=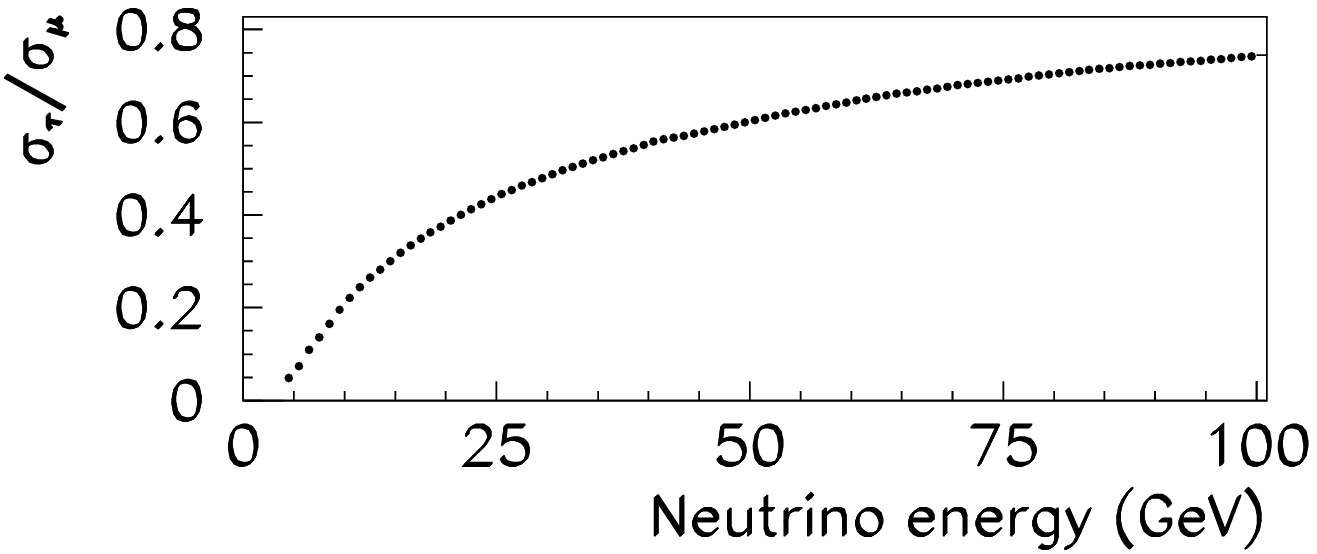,height=5cm,width=7cm}}
\end{tabular}
\vspace{2mm}
\noindent
\small
\end{center}
\label{xsec}
{\sf Figure~1:}~Left: Cross-section for \nmu{} (upper curve) and \ntau{} (lower curve) CC reactions as
function of \enu. Right: Ratio of both cross-sections, the threshold behaviour of $\sigma_{\tau}$ is
clearly visible.
\normalsize

\begin{center}

\mbox{\epsfig{file=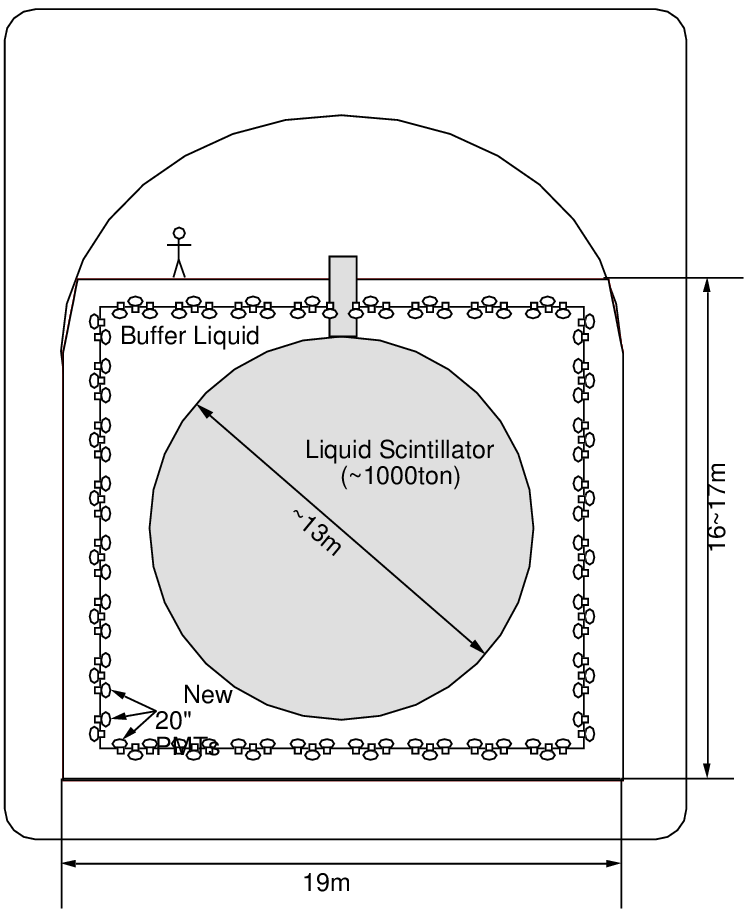,height=7cm,width=6cm}}

\vspace{2mm}
\noindent
\small
\end{center}
\label{kam}
{\sf Figure~2:}~Principle layout of the KamLAND detector currently under installation in the Kamioka mine. A plastic 
balloon filled with 1000t of Liquid Scintillator is surrounded by water as buffer liquid and a grid of 1300 20''
\pmts. 
\normalsize

\begin{center}

\mbox{\epsfig{file=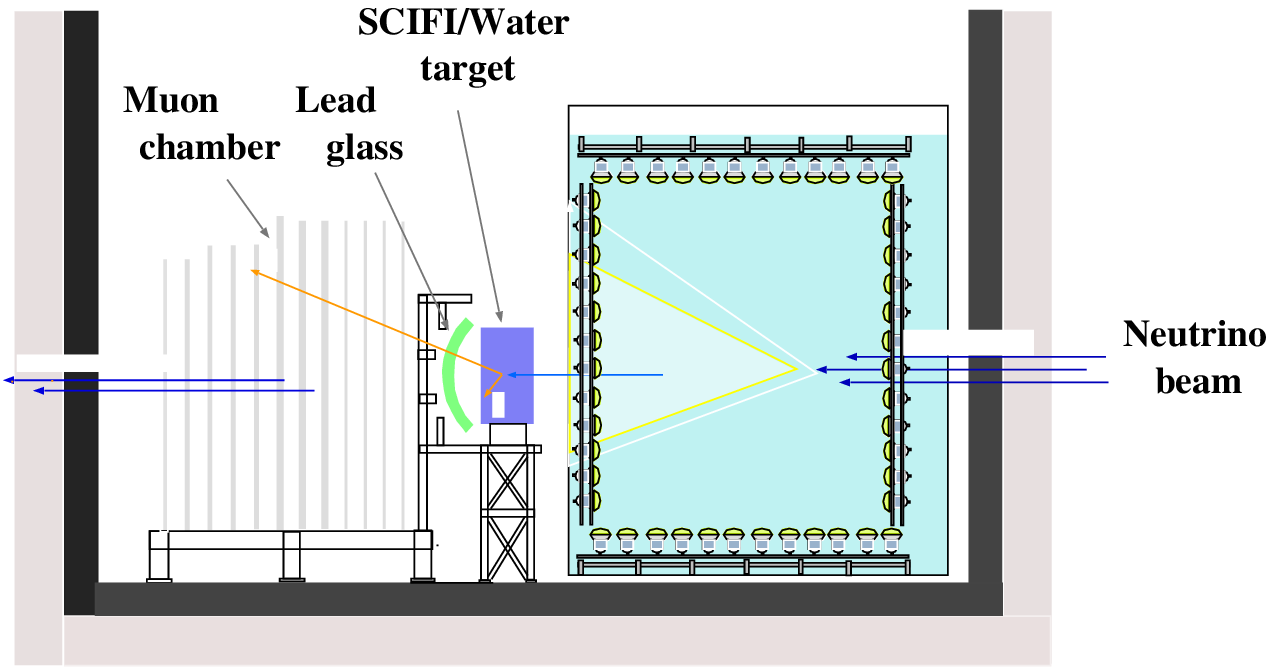,height=6cm,width=9cm}}

\vspace{2mm}
\noindent
\small
\end{center}
\label{k2knd}
{\sf Figure~3:}~Schematic design of the near detector in the K2K-experiment. The \neu{} beam is entering from the
right side, passing through a 1 kt Water-Cerenkov detector and a second fine grained detector. For details see text.
\normalsize

\begin{center}

\mbox{\epsfig{file=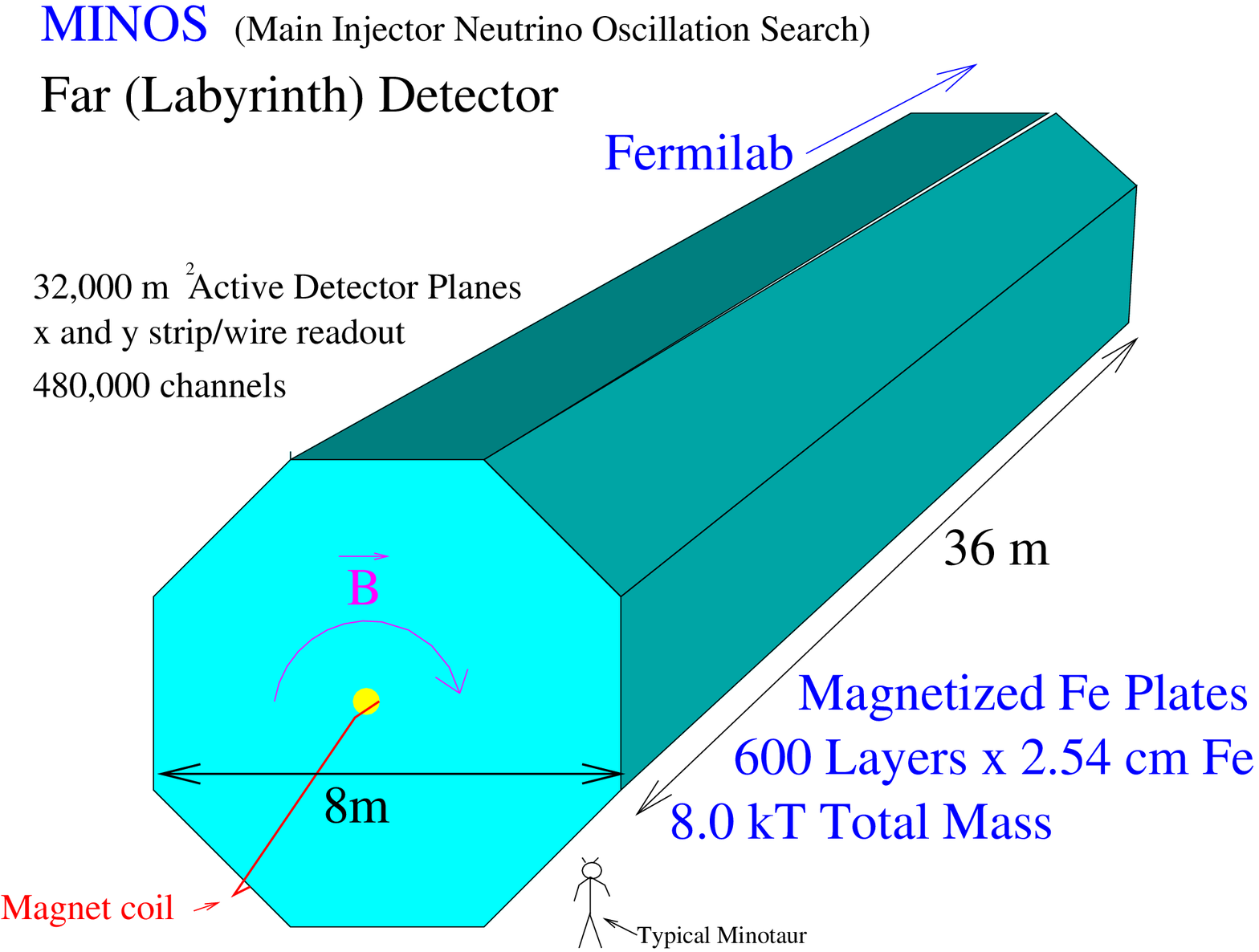,height=6cm,width=8cm}}

\vspace{2mm}
\noindent
\small
\end{center}
\label{minosdet}
{\sf Figure~4:}~Layout of the MINOS detector installed in the Soudan-mine. 
\normalsize

\begin{center}
\begin{tabular}{cc}
\mbox{\epsfig{file=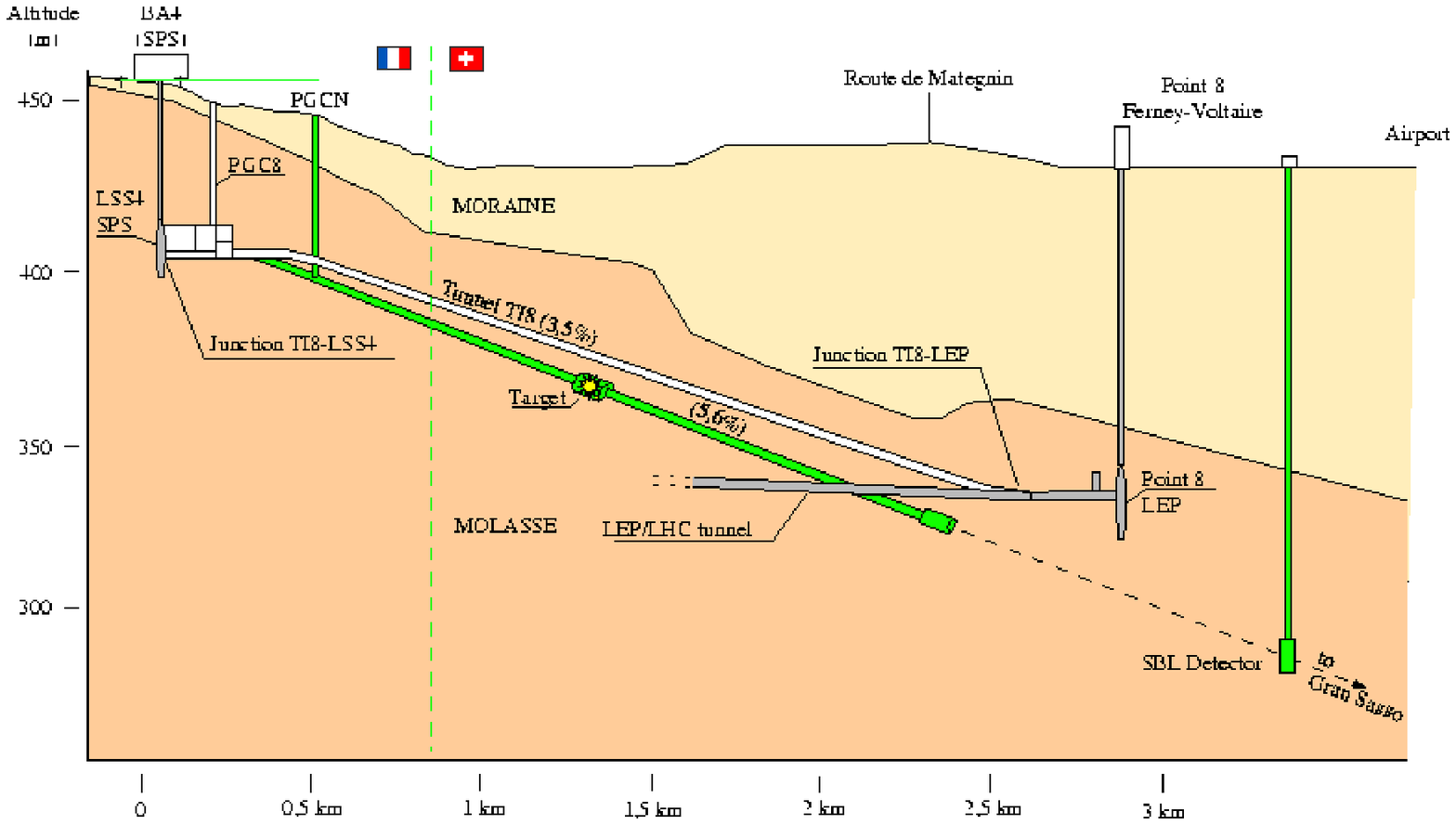,height=7cm,width=9cm}}
\mbox{\epsfig{file=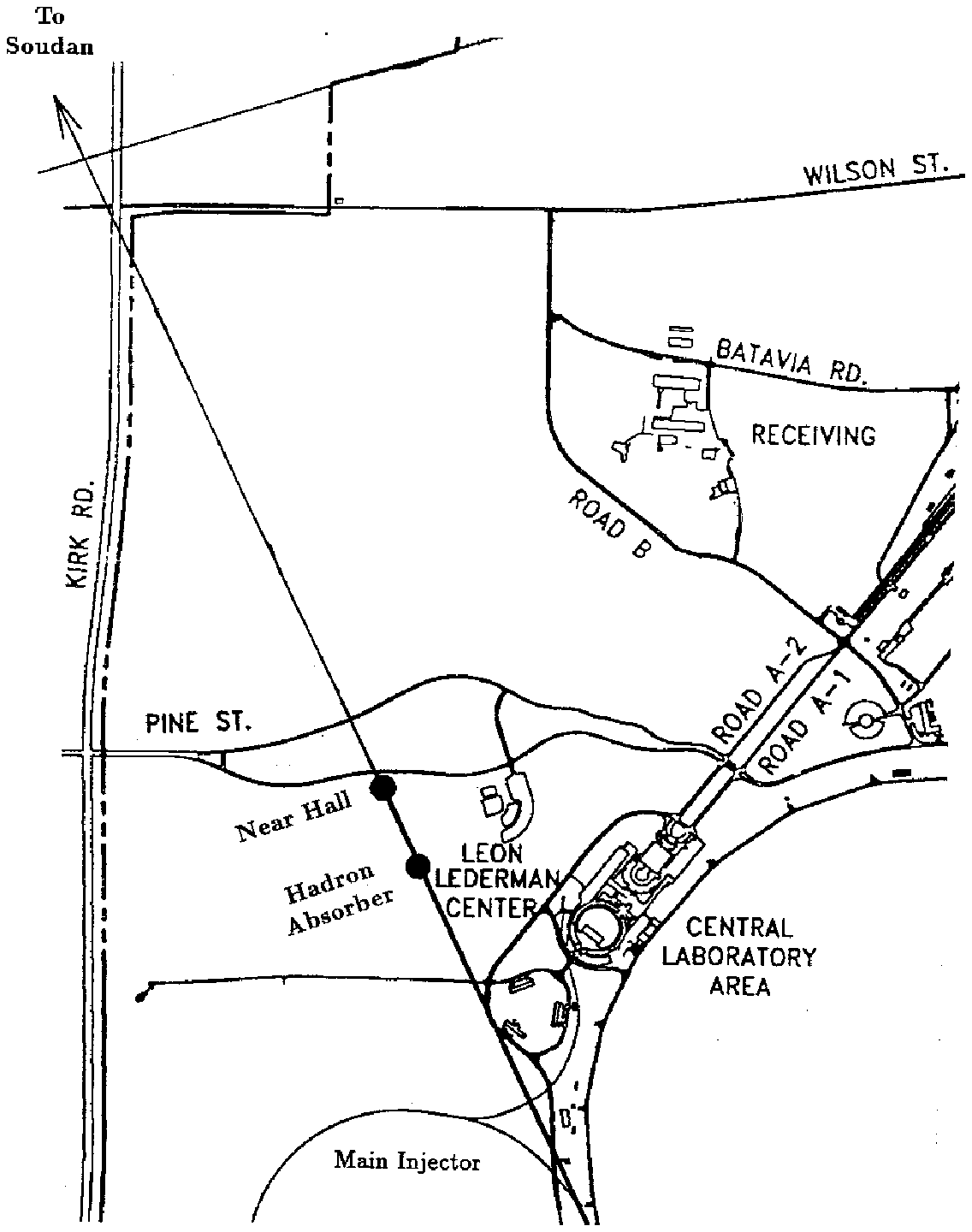,height=7cm,width=6cm}}
\end{tabular}
\vspace{2mm}
\noindent
\small
\end{center}
\label{ngs}
{\sf Figure~5:}~Left: Layout of the \neu{} beam from CERN to Gran Sasso (NGS). Right: The NuMI beam design at
Fermilab.
\normalsize

\begin{center}
\begin{tabular}{cc}
\mbox{\epsfig{file=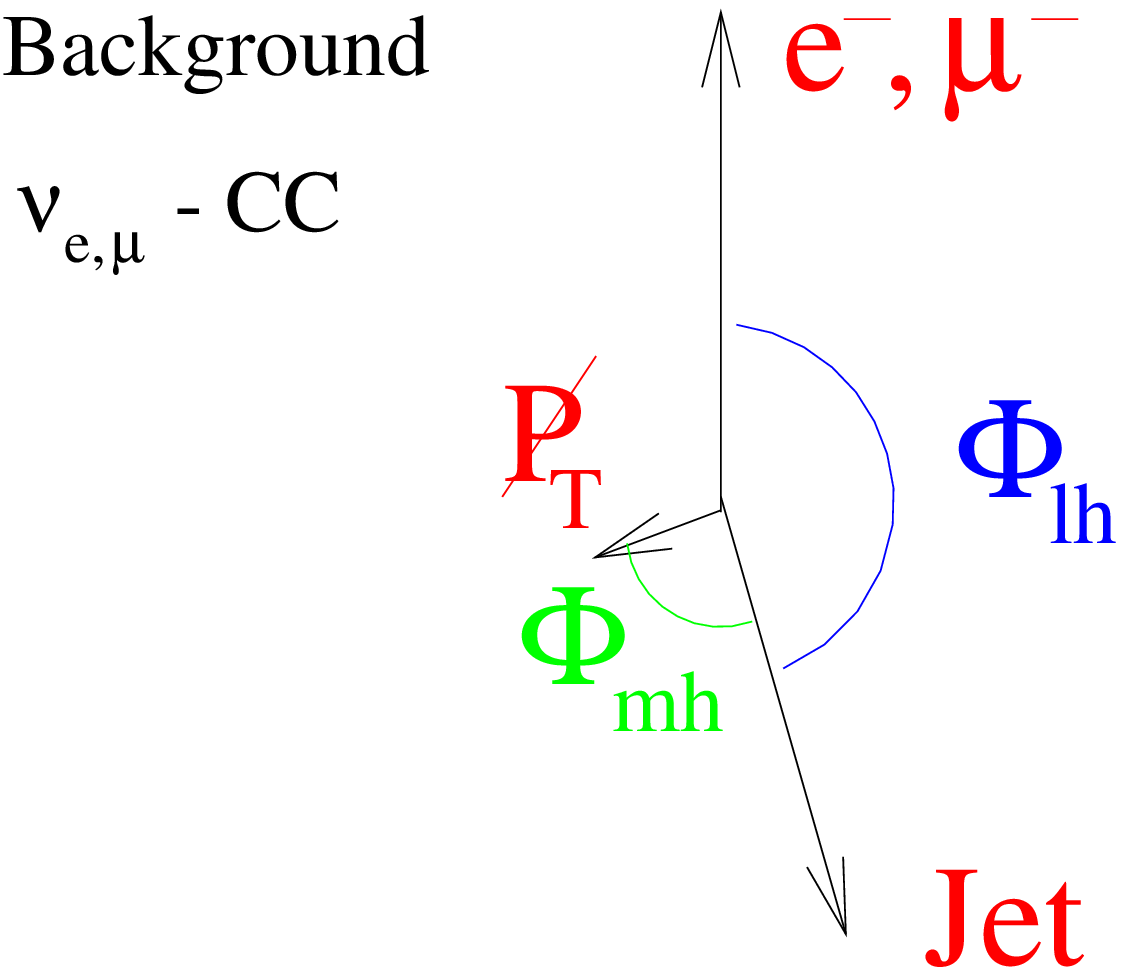,height=5cm,width=5cm}} &
\mbox{\epsfig{file=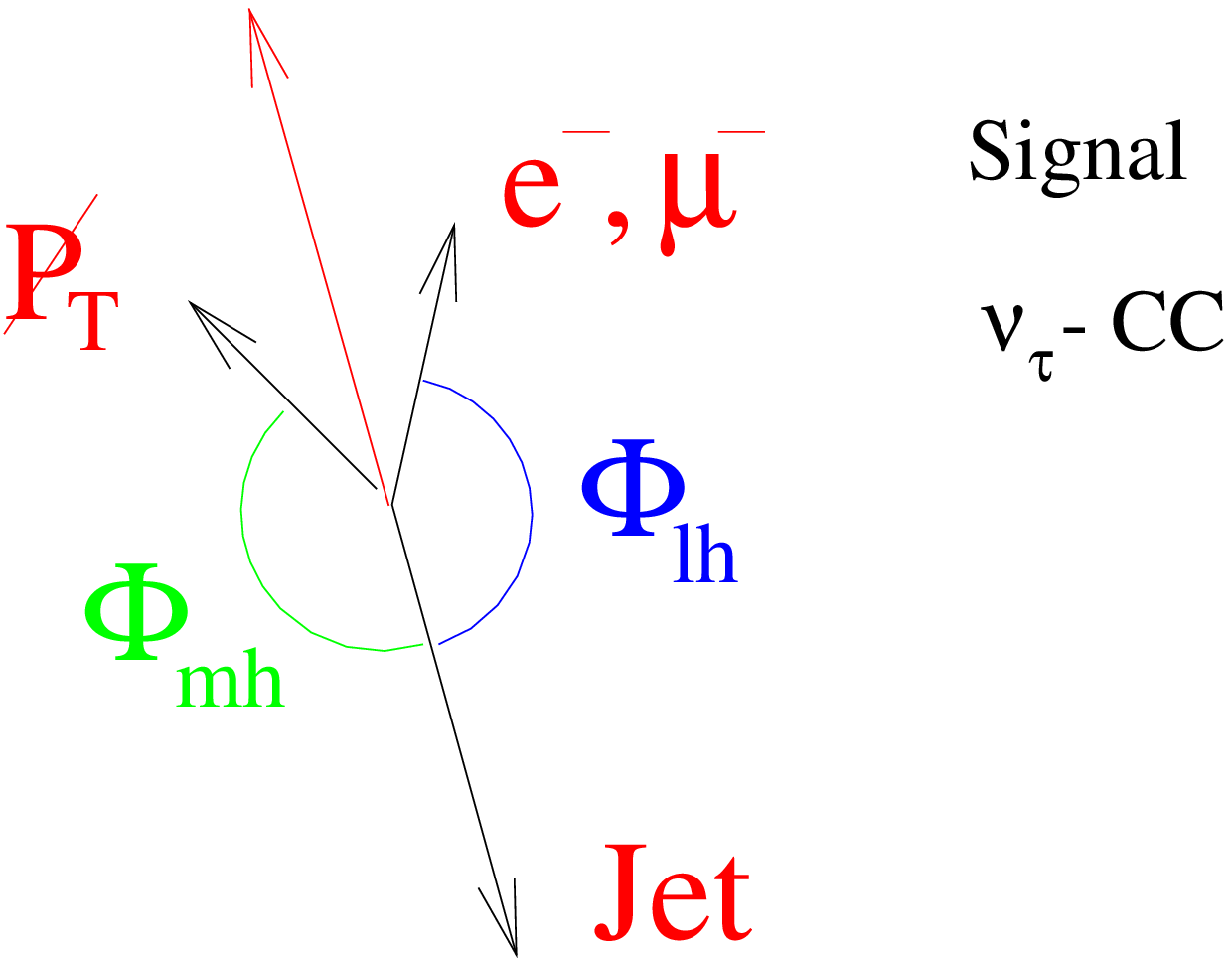,height=5cm,width=5cm}}
\end{tabular}
\vspace{2mm}
\noindent
\small
\end{center}
\label{phiphi}
{\sf Figure~6:}~Definition of the kinematic variables used in some \lbls{} for \ntau{} \app{} searches. For details
see text.
\normalsize

\begin{center}
\begin{tabular}{cc}
\mbox{\epsfig{file=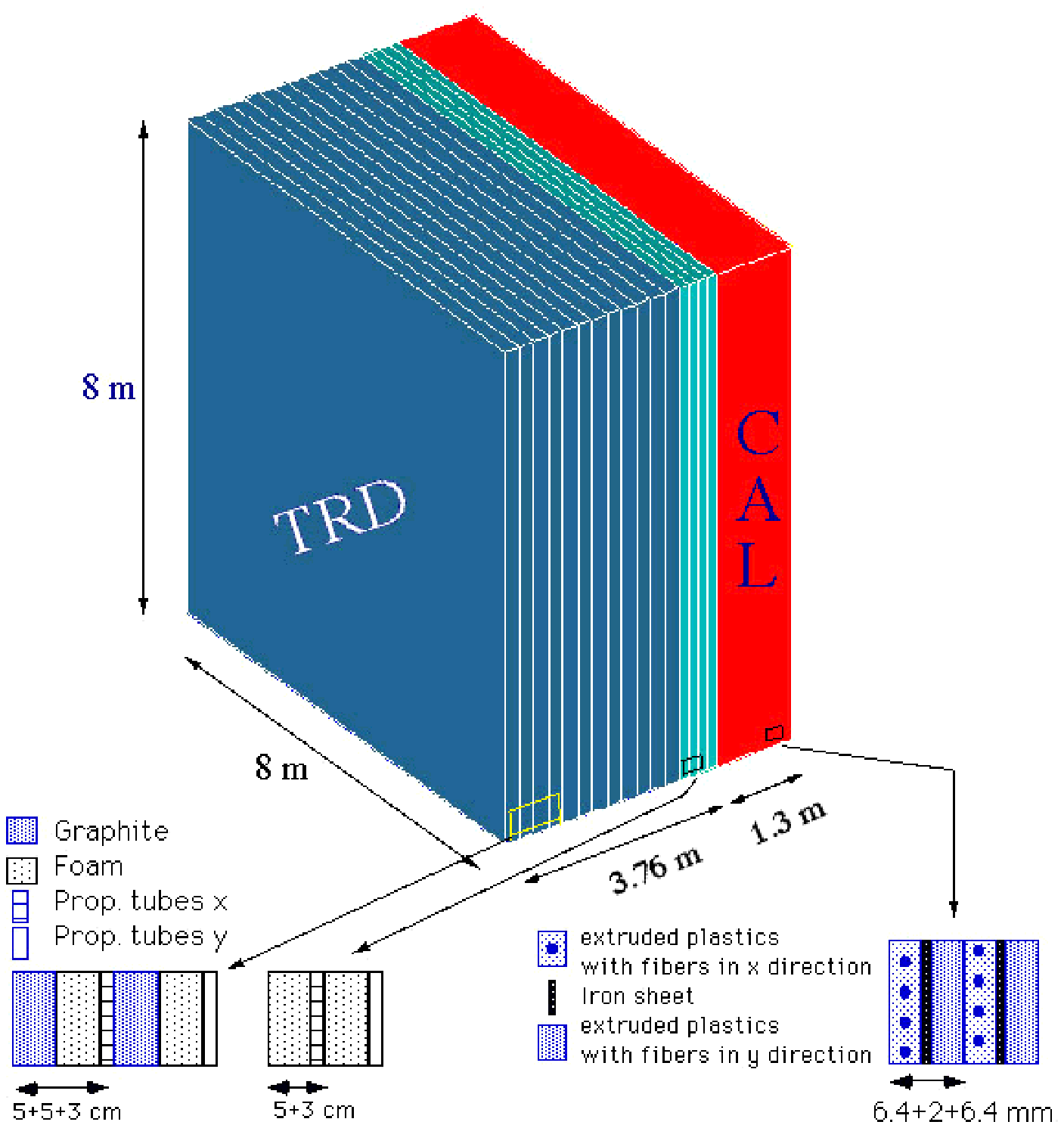,height=5.5cm,width=7cm}}
\mbox{\epsfig{file=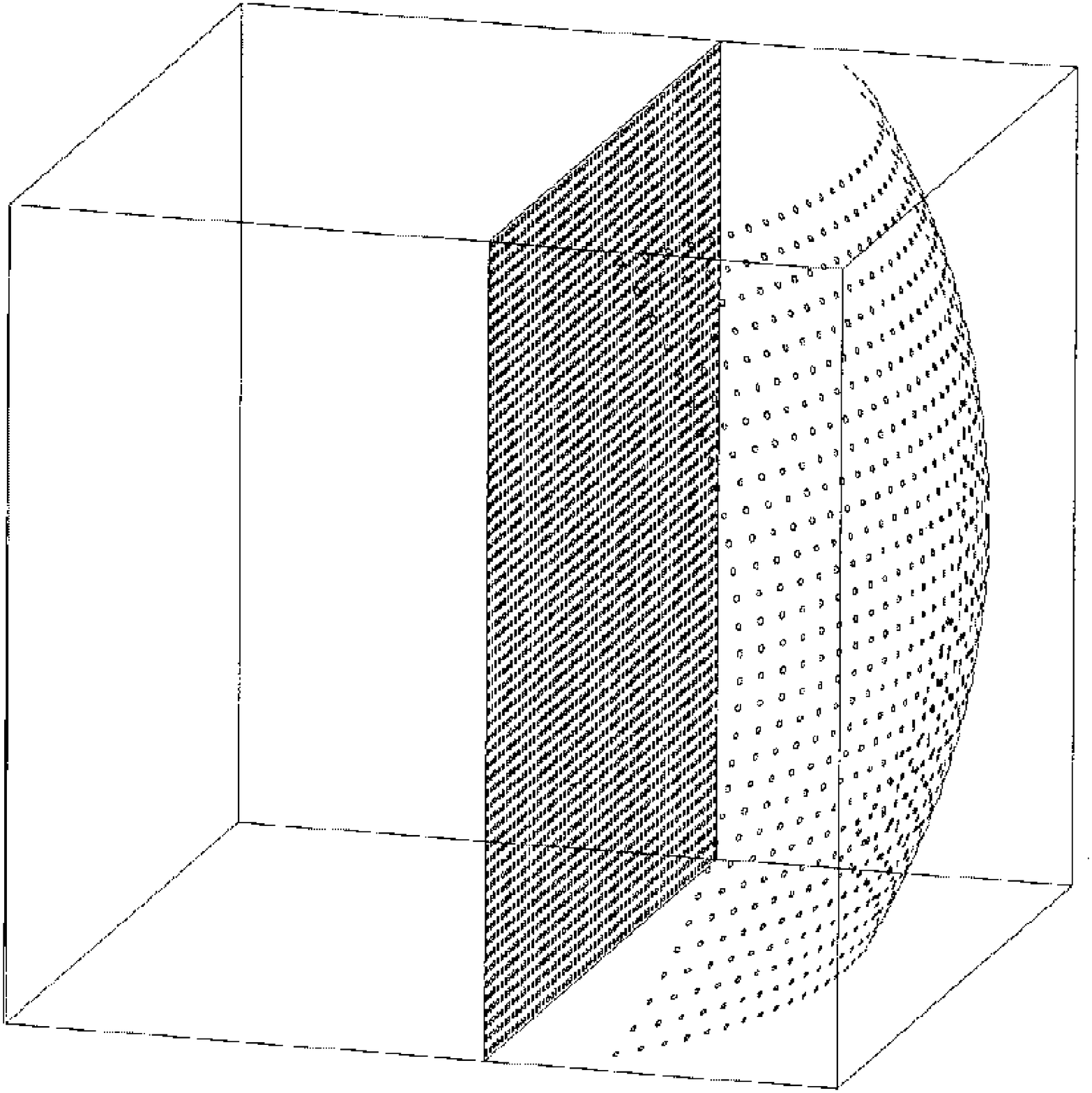,height=5.5cm,width=7cm}}
\end{tabular}

\vspace{2mm}
\noindent
\small
\end{center}
\label{noemod}
{\sf Figure~7:}~Left: Layout of the new designed NOE-module consisting of several TRDs and a calorimeter device.
Right: Principal layout of the AQUA-RICH detector. The \neu{} beam is entering a 50 m cube
containing 125 kt of water from the left side. The mirror plane (dots) and detector plane (black) are shown as well.
\normalsize

\begin{center}

\mbox{\epsfig{file=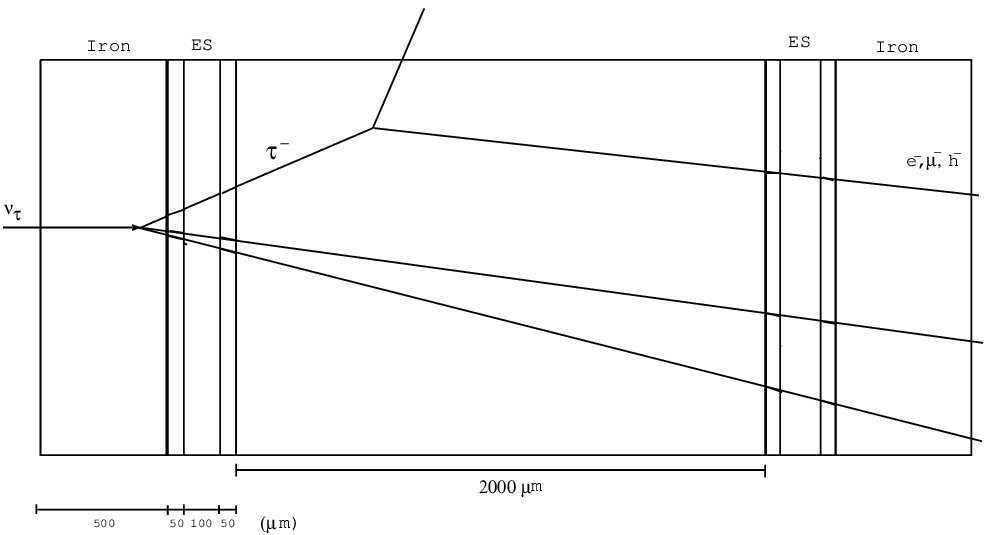,height=5.5cm,width=7cm}}

\vspace{2mm}
\noindent
\small
\end{center}
\label{opera}
{\sf Figure~8:}~Principle idea of the OPERA concept. A $\tau$-lepton produced via CC in the iron is reaching an air
gap as decay area. Using the emulsion sheets (ES) the mismatch of the track from the $\tau$ and the one from its
decay product can be seen.
\normalsize
\begin{center}

\mbox{\epsfig{file=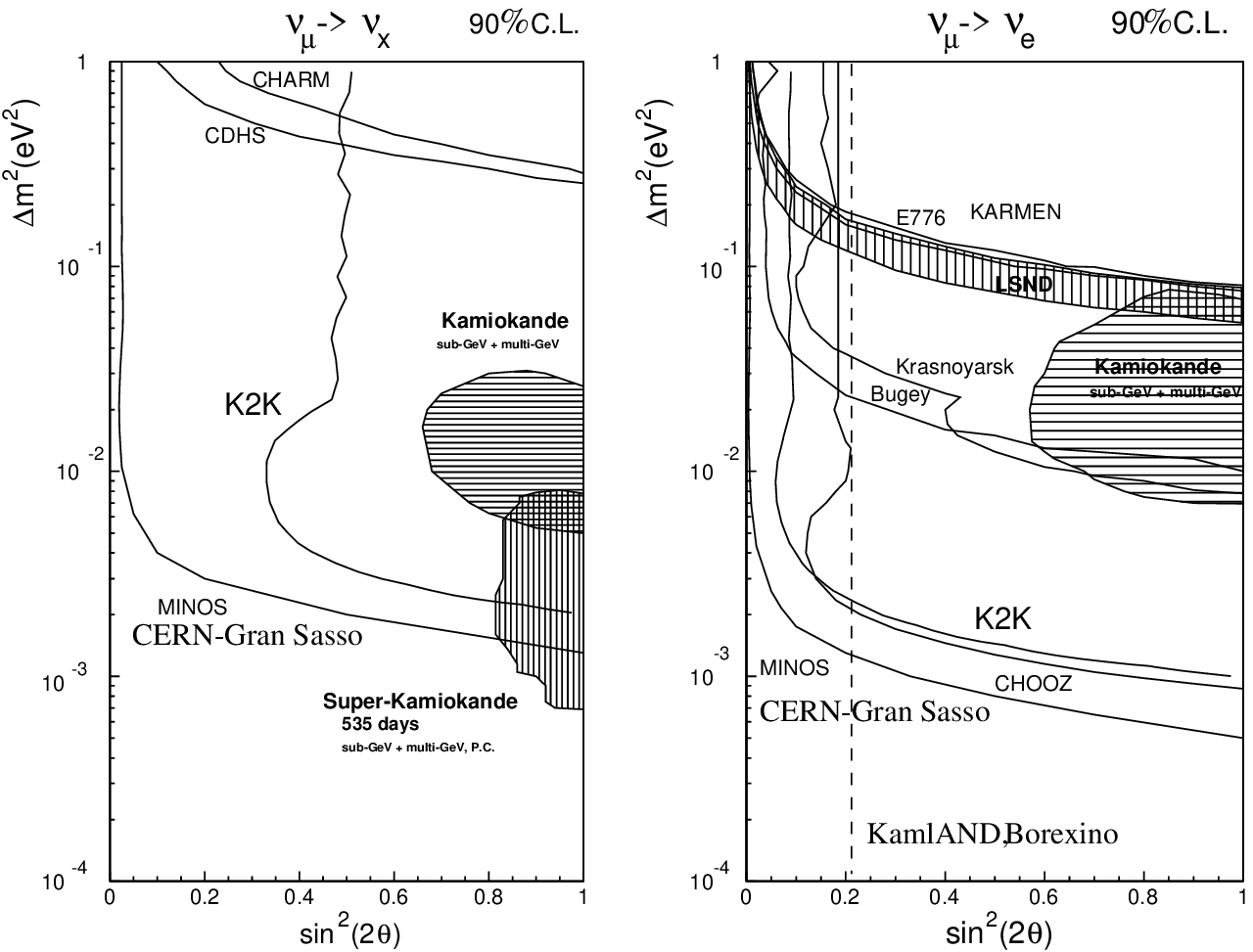,height=6cm,width=9cm}}

\vspace{2mm}
\noindent
\small
\end{center}
\label{future}
{\sf Figure~9:}~ Principle \sint{} vs. \delm{} exclusion plots of the proposed \lbls{}. Shown is the atmospheric
\neu{} region as well as given or proposed exlusion plots for $\nmu - \nu_X$ (left) and \nmune{} (right) \oszs.
\normalsize
\end{document}